\documentclass[12pt]{article}
\usepackage{latexsym, amssymb}
\def\beq{\begin{equation}}
\def\eeq{\end{equation}}
\def\beqs{\begin{equation*}}
\def\eeqs{\end{equation*}}
\def\bea{\begin{eqnarray}}
\def\eea{\end{eqnarray}}

\def\bay{\begin{array}}
\def\eay{\end{array}}
\def\beas{\begin{eqnarray*}}
\def\eeas{\end{eqnarray*}}
\def\bq{\begin{quote}}
\def\eq{\end{quote}}

\def\a{\alpha}

\def\c{\gamma}

\def\l{\lambda}

\def\s{\sigma}

\def\nn{\nonumber}

\def\t{\tilde}

\def\t2{{\textstyle{2}}}

\def\thalf{\textstyle{ {1 \over 2} } }

\def\sqr2{\sqrt{2}}

\def\and{\quad {\hbox \mathrm{and}} \quad}

\def\gappeq{\mathrel{\rlap {\raise.5ex\hbox{$>$}}
{\lower.5ex\hbox{$\sim$}}}}
\def\lappeq{\mathrel{\rlap{\raise.5ex\hbox{$<$}}
{\lower.5ex\hbox{$\sim$}}}}

\begin{document}
\pagestyle{empty}
\begin{flushright}
NMCPP/99-7\\
hep-ph/9901285\\
\end{flushright}
\vspace*{1cm}
\begin{center}
{\bf\Large Condensates Break Chiral Symmetry}\\
\vspace*{1.0cm}
{\bf Kevin Cahill\footnote{kevin@kevin.phys.unm.edu
\quad http://kevin.phys.unm.edu/$\tilde{\ }$kevin/}}\\
\vspace{.2cm}
\vspace{.2cm}
Center for Particle Physics\\
Department of Physics and Astronomy\\
University of New Mexico\\
Albuquerque, NM 87131-1156\\
\vspace{.50cm}
\vspace{.50cm}
\vspace*{1.0cm}
{\bf Abstract}
\end{center}
\begin{quote}
In the physical vacuum of $QCD$,
the energy density of light-quark fields 
strongly coupled to slowly varying gluon fields
can be negative.
The states that drive this energy density lowest
are condensates of pairs of quarks and antiquarks
of nearly opposite momenta.
These quark-antiquark condensates
break chiral symmetry. 
They may also affect
other features of hadronic physics,
such as the range of the strong force
and the confinement of color.
\end{quote}
\vfill
\begin{flushleft}
\today\\
\end{flushleft}
\eject

\pagestyle{plain}

\section{The $QCD$ Vacuum}
The thesis of this note is
that the energy density
of strongly coupled light quarks can be negative
and that this feature of $QCD$ 
breaks chiral symmetry.
The hamiltonian $H_q$ of the $u$, $d$, and $s$ quarks
\beq
H_q = 
\sum_{f=u,d,s} \int \! d^3x \, 
\bar \psi_f \left( \vec \c \cdot \vec \nabla 
- i g \c^0 A_{0a} \frac{\l_a}{2} 
- i g \vec \c \cdot \vec A_a \frac{\l_a}{2} + m_f \right) \psi_f
\label {H}
\eeq
can assume large negative mean values 
due to the term $ - g \int \! d^3x \vec J_a \cdot \vec A_a $
when the gauge field $ \vec A_a $
varies slowly with a modulus
$ |\vec A_a| $ 
that exceeds $ m_u/g $ by a sufficient margin~\cite{CahillHerling}.
For nearly constant gauge fields $ \vec A_a $,
the states that drive the energy lowest
are condensates~\cite{Wilczek} of pairs of light quarks and 
antiquarks of opposite momenta;
in such pairs the color charges cancel, but 
the color currents add.
When $ g |\vec A_a| \gg m_d $,
the $u$ and $d$ quarks play very similar roles,
and \(s \bar s \) pairs become important
when $ g |\vec A_a| \gg m_s $\@.
\par
If the gauge fields are not only slowly varying
but also essentially abelian,
in the sense that $ g f_{abc} A_\mu^b A_\nu^c $
is small 
(\emph{e.g.}, because $ A^a_\mu(x) \simeq C^a(x) V_\mu(x) $),
then the energy of the gauge fields is also small.
If an essentially abelian gauge field,
\emph{e.g.,} $|\vec A_8|$, is relatively constant 
over a sphere of radius $R$ beyond which it either
remains constant or slowly drops to zero,
then its energy density near the sphere
can be of the order of
$|\vec A_8|^2/R^2$ or less 
while that of the light-quark
condensate can be as negative as $- |g \vec A_8|^4$. 
The physical vacuum
of $QCD$ is therefore a linear combination 
of states, each of which is approximately
a coherent~\cite{Glauber} state $ | \vec A \rangle $ of 
a slowly varying, essentially abelian gauge field $ \vec A_a(x) $
and an associated condensate of pairs of
$u$ and $d$ quarks and $ \bar u $ and $ \bar d $ 
antiquarks of nearly opposite momenta:
\beq
| \Omega \rangle \simeq \int \!\!\! D\vec A_a \, f(\vec A_a)
\!\!\prod_{S(A,u)}
\!\!a^\dagger( \vec p_i, \s, u_i ) a^{c\dagger}( \vec q_j, \tau, u_j )
\!\!\prod_{S(A,d)}
\!\!a^\dagger( \vec p_i, \s, d_i ) a^{c\dagger}( \vec q_j, \tau, d_j )
| \vec A \rangle.
\label {qcd vacuum}
\eeq
Here the sets $ S(A,u) $ and $ S(A,d) $ specify
the momenta $\vec p, \vec q$, spins $\s, \tau$,
and colors $i, j$ of the quarks and antiquarks, 
the operator $ a^{c\dagger}( \vec q, \tau, d_j ) $
creates a $d$ antiquark of momentum $ \vec q $,
spin $\tau$, and color $j$,
the function $ f(\vec A_a) $ is a weight function, 
and the $s$ quarks have been suppressed.
\par
In what follows I shall compute the energy 
density of a such a light-quark condensate
for the case of a constant gauge field 
$ \vec A_8 $.  It will turn out that if
$ g |\vec A_8| $ is of the order of a GeV,
then the mean value 
$ \langle \thalf ( \bar u u + \bar d d ) \rangle $
of the light-quark condensate
is about $ (260\, $MeV$)^3$ as required
by soft-pion physics~\cite{WeinbergII}.
This condensate breaks chiral symmetry.
It may also play a role in other 
features of hadronic physics, such
as the short range of the strong force
and the confinement of quarks.

\section{A Particular Condensate}
Let us consider the case of a constant
gauge field $ \vec A_8 $ that points in the
direction 8 of color space;  
the energy density and quark condensate
associated with a slowly varying gauge 
field \( \vec A_c(\vec x) \) should be similar.
If we call the quark colors
red, green, and blue, then the condensate
will be made of red and green $u$, $d$, and $s$
quarks of momentum $ \vec p $ and both spin indices $ \s $;
red and green $u$, $d$, and $s$ antiquarks of momentum $ - \vec p $
and both spin indices $ \s $;
blue $u$, $d$, and $s$
quarks of momentum $ - \vec p $ and both spin indices $ \s $;
and blue $u$, $d$, and $s$
antiquarks of momentum $ \vec p $  
and both spin indices $ \s $.
The domains of integration $S(A,u)$ and $S(A,d)$
for the $ u $ and $d$ quarks will be very
similar when the gauge field $ \vec A_8 $ is intense,
but the domain for the $s$ quarks will be smaller.
The component $ |\Omega_A \rangle $
of the $QCD$ vacuum associated with the gauge field 
$ \vec A_8 $ will then be 
\beq
| \Omega_A \rangle = 
\prod_{S(A,u)}
a^\dagger( \vec p_i, \s, u_i ) a^{c\dagger}( - \vec p_i, \s, u_i )
\prod_{S(A,d)}
a^\dagger( \vec p_i, \s, d_i ) a^{c\dagger}( - \vec p_i, \s, d_i )
\, | \vec A_8 \rangle 
\label {cond}
\eeq
apart from the $s$ quarks.
These products over momentum, spin, and color
are defined by box quantization in a volume \(V\),
and so the mean value of the hamiltonian \( H_q \)
in the state \( | \Omega_A \rangle \) is really
an energy density.
\par
The only quark operators that have non-zero mean values
in the state $| \Omega_A \rangle $ are those
that destroy and create the same kind of quark
or antiquark.  Thus if we normally order the
quark hamiltonian (\ref{H}), then the part 
of the magnetic term 
$ H_{qm} = - g \int \!\! d^3x \, \vec J_a \cdot \vec A_a $
that involves the field $ \psi_d(x) $ of the $d$ quark 
\beq
\psi_{\ell d}(x) = \sum_\sigma \! \int \! \!\! \frac{d^3p}{(2\pi)^{3/2}}
\left[ u_\ell(\vec p, \s, d_i ) e^{ip\cdot x} a(\vec p, \s, d_i)
+v_\ell(\vec p, \s, d_i ) e^{-ip\cdot x} a^{c \dagger} (\vec p, \s, d_i)
\right]
\label {psi}
\eeq
has a mean value
\bea
E_{dm} & = & \langle \Omega_A | H_{dm} | \Omega_A \rangle = 
\langle \Omega_A | \left( - g \int \!\! d^3x 
\, \vec J_a^d \cdot \vec A_a \right) | \Omega_A \rangle \nn\\
& = & \mbox{}
\langle \Omega_A | \left( -ig \int \!\! d^3x
\, \bar \psi_d \vec \c \cdot \vec A_a \frac{\l_a}{2} \psi_d \right)
| \Omega_A \rangle
\eea
given by
\bea
E_{dm} & = & g \vec A_8 \cdot \sum_{\s,i}
\frac{\l^8_{ii}}{2}
\int_{S(A,d)} \!\!\!\!\!\!\! d^3p \, \left[
u^\dagger(\vec p_i, \s, d_i ) 
\c^0 \vec \c 
\,u(\vec p_i, \s, d_i ) \right. \nn\\
& & \mbox{} 
\left. \qquad \qquad \qquad - v^\dagger(- \vec p_i, \s, d_i )
\c^0 \vec \c 
\,v(- \vec p_i, \s, d_i ) \right].
\eea
The spin sums~\cite{Weinberg}
\beq
\sum_{\s} u_\ell(\vec p, \s, d_i ) u^\ast_{\ell'} (\vec p, \s, d_i )
= \frac{1}{2p^0} \left[ \left( p^\mu \c_\mu + i m_d\right) \c^0 
\right]_{\ell \ell'}
\label {sumu}
\eeq
and
\beq
\sum_{\s} v_\ell(\vec p, \s, d_i ) v^\ast_{\ell'} (\vec p, \s, d_i )
= \frac{1}{2p^0} \left[ \left( p^\mu \c_\mu - i m_d\right) \c^0 
\right]_{\ell \ell'}
\label {sumv}
\eeq 
imply the trace relations
\beq
\sum_{\s} u^\dagger(\vec p_i, \s, d_i )
\c^0 \vec \c
\, u(\vec p_i, \s, d_i ) = 
\sum_{\s} v^\dagger( \vec p_i, \s, d_i )
\c^0 \vec \c
\, v( \vec p_i, \s, d_i ) = - \frac{ 2 \vec p}{p^0}
\label {gamma0gamma}
\eeq
\beq
\sum_{\s} u^\dagger(\vec p_i, \s, d_i )
\c^0 
\, u(\vec p_i, \s, d_i ) = 
\sum_{\s} - v^\dagger( \vec p_i, \s, d_i )
\c^0 
\, v( \vec p_i, \s, d_i ) = - i \frac{ 2 m_d}{p^0}\!
\label {ubaru}
\eeq
and
\beq
\sum_{\s} u^\dagger(\vec p_i, \s, d_i ) \, u(\vec p_i, \s, d_i )
= \sum_{\s} v^\dagger( \vec p_i, \s, d_i ) \, v( \vec p_i, \s, d_i )
= 2. 
\label {udagu}
\eeq
It follows from the trace relation (\ref{gamma0gamma})
that the magnetic energy density of the $d$ quarks
and antiquarks of color $i$
in the constant gauge field $ \vec A_8 $ is
\beq
E_{dm} = - 2 g \l^8_{ii} \int_{S(A,d)} \!\! 
\frac{ d^3p }{  p^0} \,
\vec A_8 \cdot \vec p .
\eeq
The trace relation (\ref{udagu}) implies that 
in the state $ | \Omega_A \rangle $  
the mean value of the color charge density 
$ J_a^{0d} = \psi_d^\dagger \thalf \l_a \psi_d $
and therefore that of the second term of the hamiltonian (\ref{H}) 
vanish.
It follows from the Gell-Mann matrices \( \l_a \)
and from the trace relation (\ref{gamma0gamma}) that the color current 
$ \vec J_a^d = - \psi_d^\dagger \c^0 \vec \c \thalf \l_a \psi_d $ 
for $ a \ne 8$ also vanishes.
\par
The mean value of the hamiltonian \( H_q \)
for $d$ quarks and antiquarks of color $i$
in the state \( | \Omega_A \rangle \)
is therefore
\beq
E_{{di}} = \int_{S(A,d,i)} \!\!  d^3p  
\left(
4 p^0 - 2 g \l^8_{ii} \frac{\vec A_8 \cdot \vec p}{ p^0} 
\right)
\label {Edi}
\eeq   
where the domain of integration $S(\vec A_8,d,i) $
is the set of momenta $ \vec p $ for which the integrand is negative
\beq
g \l^8_{ii} \vec A_8 \cdot \vec p > 2 ( \vec p^2 +m_d^2 ) .
\label {domain}
\eeq
The set $S(\vec A_8,d,i) $ is empty unless 
\( g \l^8_{ii} |\vec A_8| > 4 m_d \), 
which requires
the effective magnitude of the gauge field 
to be large compared to the mass $ m_d $ of the $d$ quark.
\par
Since the quark energy density \( E_{fi} \) 
depends upon the flavor $f$ and the color $i$ only through
the dimensionless ratio
\beq
r = \frac{4 m_f }{g \l^8_{ii} |\vec A_8|},
\eeq
we may write it as the integral 
\beq
E_{fi} = - 32 \pi (g \l^8_{ii} |\vec A_8|)^4
\int_{1-\sqrt{1-r^2}}^{1+\sqrt{1-r^2}}
dx \, \frac{x ( 2x - x^2 - r^2 )^2 }{ \sqrt{ x^2 + r^2 }}
\eeq
which has the value
\bea
E_{fi} & = & \mbox{} - 32 \pi (g \l^8_{ii} |\vec A_8|)^4 \nn\\ 
& & \mbox{} \left[
\left( \frac{1}{5} (x^2+r^2)^2
- x^3 - \frac{1}{2} r^2 x
+ \frac{\textstyle 4}{\textstyle 3} (x^2 - 2 r^2 )
\right)
\sqrt{x^2 + r^2} \right. \nn\\
& & \left. \qquad \qquad 
\mbox{} + \frac{1}{2} r^4 {\rm arcsinh}{\left(\frac{x}{r}\right)}
\right]_{1-\sqrt{1-r^2}}^{1+\sqrt{1-r^2}} .
\eea
For small $r$ the energy density $E_{fi}$ is approximately
\beq
E_{fi} \simeq - 32 \pi (g \l^8_{ii} |\vec A_8|)^4
\left( \frac{16}{15} 
- 4 \left(\frac{4 m_f}{g \l^8_{ii} |\vec A_8|} \right)^2 \right).
\eeq
Summing over the three colors, we get
\beq
E_{f} \simeq - 64 \pi (g |\vec A_8|)^4
\left( \frac{16}{15}
- 4 \left(\frac{4 m_f}{g |\vec A_8|} \right)^2 \right)
\eeq 
which displays isospin symmetry 
when \( 4 m_f \ll g | \vec A_8 | \).
If the gauge field is moderately strong
\( 4 m_s > g \l^8_{ii} | \vec A^8 | \gg 4 m_d \),
then the energy density of the $u$--$d$ condensate is
\beq
E_{ud} \simeq - 128 \pi (g |\vec A_8|)^4
\left( \frac{16}{15}
- 4 \left(\frac{4 m_{ud}}{g |\vec A_8|} \right)^2 \right)
\eeq 
where \( m_{ud}^2 = ( m_u^2 + m_d^2 )/2 \).
For stronger gauge fields,
\( g \l^8_{ii} | \vec A^8 | >> 4 m_s \),
the energy density of the light-quark condensate is
\beq
E_{uds} \simeq - 192 \pi (g |\vec A_8|)^4
\left( \frac{16}{15}
- 4 \left(\frac{4 m_\ell}{g |\vec A_8|} \right)^2 \right)
\eeq 
where \( m_\ell^2 = ( m_u^2 + m_d^2 + m_s^2 )/3 \).

\section{The Breakdown of Chiral Symmetry}
The quark condensate occasioned by the constant
gauge field $ \vec A_8 $ gives rise to a mean value
of the space average of $ \thalf ( \bar u u + \bar d d ) $,
which is an order parameter that traces
the breakdown of chiral symmetry.
It follows from the expansion (\ref{psi}) of the Dirac field
and the trace relation (\ref{ubaru}) 
that this order parameter is
\bea
\langle {\thalf} ( \bar u u + \bar d d ) \rangle 
& = & \langle \Omega_A | \int \!\! d^3x {\thalf} ( \bar u u + \bar d d )
| \Omega_A \rangle 
\nn \\ & = & 
{\thalf} \sum_{f=u}^d \sum_{i=1}^3
\sum_\s \int_{S(A,f,i)} \!\! d^3p \left[
u^\dagger(\vec p_i, \s, f_i ) i \c^0  
u(\vec p_i, \s, f_i ) \right.
\nn \\ & & \mbox{}
\left. \qquad \qquad \qquad \qquad
- v^\dagger(- \vec p_i, \s, f_i ) i \c^0 
v(- \vec p_i, \s, d_i ) \right]
\nn \\ & = & \mbox{}
\sum_{f=u}^d \sum_{i=1}^3   
2 m_f \int_{S(A,f,i)} \!\! \frac {d^3p}{p^0} .
\eea
In terms of the ratio \( r = 4 m_f/(g \l^8_{ii} | \vec A_8 |) \),
it is
\bea
\langle {\thalf} ( \bar u u + \bar d d ) \rangle 
\!\!& = & \!\!\! \sum_{f=u}^d \! \sum_{i=1}^3
\frac{\pi}{4} m_f \left( g \l^8_{ii} | \vec A_8 | \right)^2
\! \int_{1-\sqrt{1-r^2}}^{1+\sqrt{1-r^2}} 
\!\!\!\!\!\!\!\!\!\! dx
\left( \frac{x^2}{\sqrt{x^2 + r^2}} 
- \frac{x}{2} \sqrt{x^2 + r^2}
\right) \nn\\
\!\!& = & \mbox{}
\!\!\! \sum_{f=u}^d \sum_{i=1}^3
\frac{\pi}{4} m_f \left( g \l^8_{ii} | \vec A_8 | \right)^2
\nn\\
& & 
\left[ \left( \frac{x}{2} - \frac{x^2+r^2}{6} \right) \sqrt{x^2+r^2} 
- \frac{r^2}{2} {\rm arcsinh} \left( \frac{x}{r} \right)
\right]_{1-\sqrt{1-r^2}}^{1+\sqrt{1-r^2}}.
\eea
\par
For small $r$ this condensate or order parameter is
\beq
\langle {\thalf} ( \bar u u + \bar d d ) \rangle
\simeq \sum_{f=u}^d \sum_{i=1}^3 
\frac{\pi}{4} m_f 
\left( g \l^8_{ii} | \vec A_8 | \right)^2 
\left[ \frac{2}{3} + 
\frac{r^2}{2} \left( \ln \left( \frac{r}{4} \right) 
- \frac{1}{2} \right) \right].
\eeq
In the limit \( r \to 0 \) and
summed over colors (and over $u$ and $d$), it is
\beq
\langle {\thalf} ( \bar u u + \bar d d ) \rangle
\simeq \frac{\pi}{3} ( m_u + m_d )
\left( g | \vec A_8 | \right)^2.
\eeq
\par
We may use this formula
and Weinberg's relation
(Eq.(19.4.46) of \cite{WeinbergII})
\beq
\langle {\thalf} ( \bar u u + \bar d d ) \rangle
\simeq \frac{m_\pi^2 F_\pi^2}{4(m_u + m_d )}
\eeq
in which \( F_\pi \simeq 184 \, \)MeV 
is the pion decay constant,
to estimate the effective magnitude \( \langle g | \vec A_8 | \rangle \)
of the gauge field in the physical vacuum of \emph{QCD} as
\beq
\langle g | \vec A_8 | \rangle \simeq
\sqrt{ \frac{3}{4\pi} }
\, \frac{ m_\pi F_\pi}{( m_u + m_d )}.
\label {gA}
\eeq
In the \(\overline{\rm MS}\) scheme 
at a renormalization scale \( \mu = 1 \,\)GeV,
the mass range
\( 3 \, \)MeV \( < ( m_u + m_d )/2 < 8 \, \)MeV
of the Particle Data Group~\cite{PDG}
implies that
\( 770 \, \)MeV \( < \langle g | \vec A_8 | \rangle < 2070 \, \)MeV\@.
Since \( \mu = 1 \,\)GeV is somewhat high
for hadronic physics, the low end of the range,
\( \langle g | \vec A_8 | \rangle \simeq 800 \, \)MeV,
may be more reliable.  

\section{Exact Ground State}
Because the hamiltonian (\ref{H})
is quadratic in the quark fields,
it is possible to find its exact ground state
when the gluon fields are replaced by their mean values.  
For the case in which the gluons are
represented by a coherent state \( | A_c^3 \rangle \)
that is constant and points in the 
z-direction \( \vec A_c = A_c^3 \)
with \( A_c^0 = 0 \)
and \( c = 3 \) or 8,
the exact ground state \( | \Omega_A^e \rangle \) 
may be defined in terms of the quantity
\beq
\a_i = {\thalf} g A_c^3 \l_{ii}^c
\label {a}
\eeq
as a product over all momenta \( \vec p \),
colors \(i\), and flavors \(f\)
\beq
| \Omega_A^e \rangle
= \left( \prod_{ \vec p, i, f }
C( \vec p, i, f, A_c^3 ) \right) | A_c^3 \rangle 
\label {eprod}
\eeq
in which the operator \( C( \vec p, i, f, A_c^3 ) \) is
\bea
\lefteqn{C( \vec p, i, f, A_c^3 ) = \frac{1}{2 p^0 s} 
\Bigl\{
\! \left( p_0^2 \! - \! \a_i p_3 \! - \! p^0 s \right)
a^\dagger_{if}(\vec p,{\thalf}) a^{c\dagger}_{if}(-\vec p,{\thalf})
a^\dagger_{if}(\vec p,-{\thalf}) a^{c\dagger}_{if}(-\vec p,-{\thalf})}
\nn\\ 
& & \mbox{} \!\!\!\!\!\! + \frac{ \a_i p_3 }{ p^0 + m_f }
\left[ \left( p_1 \! - \! i p_2 \right) 
a^\dagger_{if}(\vec p,{\thalf})
a^{c\dagger}_{if}(-\vec p,{\thalf}) 
- \left( p_1 \! + \! i p_2 \right)
a^\dagger_{if}(\vec p,-{\thalf})
a^{c\dagger}_{if}(-\vec p,-{\thalf})
\right] \nn\\
& & \mbox{} \!\!\!\!\!\!
+ \frac{ \a_i \left( p_0^2 + p^0 m_f - p_3^2 \right)}{p^0 + m_f}
\left[
a^\dagger_{if}(\vec p,{\thalf})
a^{c\dagger}_{if}(-\vec p,-{\thalf})
+ a^\dagger_{if}(\vec p,-{\thalf})
a^{c\dagger}_{if}(-\vec p,{\thalf})
\right] \nn\\
& & \mbox{} \!\!\!\!\!\!
+ \left( p_0^2 - \a_i p_3 + p^0 s \right) \Bigr\}
\label {exact}
\eea
where \( a^{c\dagger}_{if}(-\vec p,{\thalf}) =
a^{c\dagger}(-\vec p,{\thalf},i,f) \)
creates antiquarks of color \(i\) 
and flavor \(f\) and \(s\) is the square root
\beq
s = \sqrt{p_0^2 - 2 \a_i p_3 + \a_i^2} \ge 0.
\label {s}
\eeq
\par
The state \( | \Omega_A^e \rangle \) 
is an eigenstate of the hamiltonian (\ref{H})
with eigenvalue
\beq
E( A_c^3) = \sum_{i,f} \int \! d^3p \, \left( 2 p^0
- 2 \frac{\a_ip_3}{p^0} - 2 s \right) 
\label {E}
\eeq
in which the integrand \( E(\vec p, i, f, A_c^3) = 
2 p^0 - 2 \a_i p_3/p^0 - 2s \) is negative definite.
When \( \a_i p_3 \gg p_0^2 \),
which is a generous version of the inequality (\ref{domain}), 
this integrand is approximately
\beq
E(\vec p, i, f, A_c^3) \simeq 4 p^0 - 4 \frac{\a_i p_3}{p^0},
\label {4e}
\eeq
which is that of the integral (\ref{Edi}).
But when \( p_0^2 \gg \a_i p_3 \)
and \(  p_0^2 \gg \a_i^2 \), it is 
\beq
E(\vec p, i, f, A_c^3) \simeq - \a_i^2 \,
\frac{m_f^2 +p_1^2 +p_2^2}{\left( p^0 \right)^3}.
\label {aa}
\eeq
In terms of the cutoff \( \Lambda \), 
the quadratically divergent energy density (\ref{E}) is 
\beq
E( A_c^3) \sim - \left( A_c^3 \right)^2 \Lambda^2.
\label {EL}
\eeq
It exerts a force, \( \sim 2 A_c^3 \Lambda^2 \),
on the mean value of the gauge field \( A_c^3 \) driving it
to larger values.

\section{Conclusions and Speculations}
We have seen that a vacuum component
\( | \Omega_A \rangle \)
consisting of a coherent state
of a slowly varying gauge field \( \vec A_c \)
and an associated quark-antiquark condensate
(\ref{cond})
possesses a negative energy density \( \rho_\Omega \)
of the order of \(  - (g | \vec A_c |)^4 \).
The \( q \bar q \) condensate 
in the component \( | \Omega_A \rangle \)
breaks chiral symmetry with an order parameter 
\( \langle {\thalf} ( \bar u u + \bar d d ) \rangle \)
that agrees with soft-pion physics if the effective
strength \( g | \vec A_c | \) of the gauge field
is of the order of 800 MeV.

The real vacuum is an integral over 
all such components \( |\Omega_A \rangle \)\@.
In the temporal gauge, the gauge field \( A_a^0 \)
is absent, and the vacuum is an 
integral (\ref{qcd vacuum}) over all gauge transformations
\( \omega(\vec x) \)
of the image \( | \Omega_A^\omega \rangle \)
of a state like the component \( | \Omega_A \rangle \)
under the gauge transformation \( \omega(\vec x) \).
This integral removes any breaking of rotational
invariance associated with the uniform field \( \vec A_8 \).
The mean value of any gauge-invariant operator,
for instance the quark hamiltonian \( H_q \),
is a double integral
\beq
\langle \Omega | H_q | \Omega \rangle
= \int \! D \vec A_a \int \! D \vec A_a^\prime 
\, \langle \Omega_A | H_q | \Omega_{A^\prime} \rangle 
\label {mean value}
\eeq
in which most of the off-diagonal terms are very small.  

Of course, the actual energy density
of the vacuum is small and non-negative.
But by using normal ordering,
we have been ignoring the zero-point energies 
of the fields.
Zero-point energies augment \( \rho_\Omega \)
by a positive or negative 
energy density that is quartically divergent
unless the number of Fermi fields
is equal to the number of Bose fields
and is quadratically divergent unless the supertrace
\( \sum (-1)^{2j} m^2_j \)
of the squared masses of all particles vanishes.
The large negative energy density \( \rho_\Omega \)
may make it possible to cancel a large positive
energy density due to the breaking of supersymmetry.

The condensates described in this paper may explain
why massless gluons give rise to a short-range force. 
To see this let us specialize to the component \( | \Omega_A \rangle \)
of the \emph{QCD} vacuum
that has a constant gauge field \(A_8^1\)
pointing in direction 1 of space 
and direction 8 of color space.
In this component
the mean value of the hamiltonian of the gauge fields 
contains a mass term for the fields \( A_b^m \)
for \(m \ne 1 \) 
\beq
\sum_{a,b,c,m} {\thalf} f_{a8b} f_{a8c} \langle A_8^1 \rangle^2
A_b^m A_c^m
= \sum_{b,c,m} {\thalf} M_{bc}^2 A_b^m A_c^m
\eeq
in which the mass matrix is
\beq
M_{bc}^2 =
\langle g A_8^1 \rangle^2 
\sum_a f_{a8b} f_{a8c} 
= \langle g A_8^1 \rangle^2 
\left( T_8^2 \right)_{bc} 
\eeq
where the \(8\times8\) matrix \(T_8\)
is a generator 
of the group \emph{SU(3)} in the adjoint representation.
Since the matrix \(T_8\) is hermitian, 
the eigenvalues of the mass matrix
\( M_{bc}^2 \) are all non-negative.

The \emph{QCD} vacuum (\ref {qcd vacuum})
is an integral over slowly varying gauge
fields and their correlated condensates.
In this vacuum \( | \Omega \rangle \),
every space component \(A_c^i\) 
of the gauge field acquires a non-zero mean value. 
Thus gluons are massive in the vacuum of \emph{QCD},
and according to the estimate (\ref{gA}),
the mass of the gluon is in the range of hundreds of MeV. 

The condensates of this paper may also explain
quark confinement~\cite{Gribov,Grady88}.
Because a color-electric field moves 
quarks in one direction and antiquarks
in the opposite direction,
the quark condensate of 
the $QCD$ vacuum in this model
is not stable in the presence 
of color-electric fields.
Thus volumes of space that are traversed
by color-electric fields have less quark-antiquark condensate~\cite{Faber} 
and so a higher energy density than that of the physical vacuum.
Hence the surface of a hadron 
is exposed to a pressure that is equal
to the difference between the energy density 
$ \rho_\Omega $\@  
of the physical vacuum outside the hadron
and the energy density \( \rho_h \) inside the hadron
which, due to the color-electric fields,
is somewhat higher than $ \rho_\Omega $\@.
This pressure \( p \)
\beq
p \simeq \rho_h - \rho_\Omega 
\eeq
confines quarks because
it squeezes their color-electric fields.
In this view quarks are confined not because
of the energy of their color-electric fields
but because their color-electric fields
are excluded by the physical vacuum.
With the estimate (\ref{gA})
of \( \langle g | \vec A_8 | \rangle \),
a drop of only 1\% in the hadronic quark-antiquark
condensate would produce a confining pressure
of the order of \(p \simeq (1 {\rm GeV})^4 \).

The standard picture of quark confinement 
rests upon computer simulations
of lattice gauge theories.  In these theories
the gauge fields are replaced by the elements
of a compact gauge group.  This compactification
generates unwanted effects known as artifacts.
Most~\cite{ba} if not all~\cite{Grady98}
of the string tension measured in compact lattice
gauge theory is due to such artifacts. 
In non-compact lattice simulations~\cite{palumbo90},
there is no sign of confinement
unless the action contains auxiliary fields,
which introduce other kinds of artifacts.
Thus in all lattice simulations whether compact or non-compact,
confinement has appeared only
when accompanied by significant lattice artifacts.
The artifacts that drive confinement on the lattice
may represent the long-distance effects of a true,
but unknown, microscopic lattice action.
But it is also possible that these artifacts
are merely artifacts, have nothing to do with confinement,
and do not survive in the continuum limit~\cite{Grady98}.

\section*{Acknowledgements}
I should like to thank L.~Krauss
for a discussion of chiral symmetry,
S.~Weinberg for e-mail about symmetry breaking,
M.~Gell-Mann for pointing out the diquark-condensate papers
of Wilczek {\it et al.\/},
and H.~Bryant, M.~Gold,
G.~Herling, M.~Price, and G.~Stephenson
for helpful comments.

\end{document}